\documentclass[10pt,conference]{IEEEtran}
\IEEEoverridecommandlockouts

\usepackage{listings}
\usepackage{cite}
\usepackage{amsmath,amssymb,amsfonts}
\usepackage{graphicx}% http://ctan.org/pkg/graphicx
\usepackage{textcomp}
\usepackage{xcolor}
\usepackage{subcaption}
\usepackage[linesnumbered,lined,boxed,commentsnumbered]{algorithm2e}
\usepackage{algpseudocode}
\usepackage{nicematrix}
\usepackage[]{mdframed}
\usepackage{multirow}
\def\BibTeX{{\rm B\kern-.05em{\sc i\kern-.025em b}\kern-.08em
    T\kern-.1667em\lower.7ex\hbox{E}\kern-.125emX}}

\definecolor{codegreen}{rgb}{0,0.6,0}
\definecolor{codegray}{rgb}{0.5,0.5,0.5}
\definecolor{codepurple}{rgb}{0.58,0,0.82}
\definecolor{backcolour}{rgb}{0.95,0.95,0.92}

\lstdefinestyle{mystyle}{
    backgroundcolor=\color{backcolour},   
    commentstyle=\color{codegreen},
    keywordstyle=\color{magenta},
    numberstyle=\tiny\color{codegray},
    stringstyle=\color{codepurple},
    basicstyle=\ttfamily\footnotesize,
    breakatwhitespace=false,         
    breaklines=true,                 
    captionpos=b,                    
    keepspaces=true,                 
    numbers=left,                    
    numbersep=5pt,                  
    showspaces=false,                
    showstringspaces=false,
    showtabs=false,                  
    tabsize=2
}

\begin{document}

\title{Automated User Story Generation with Test Case Specification Using Large Language Model
}

\author{
    \IEEEauthorblockN{1\textsuperscript{st} Tajmilur Rahman}
    \IEEEauthorblockA{\textit{Computer Science} \\
    \textit{University of Saskatchewan}\\
    Saskatoon, SK, Canada \\
    qoy860@usask.ca}
    \and
    \IEEEauthorblockN{2\textsuperscript{nd} Yuecai Zhu}
    \IEEEauthorblockA{\textit{Enterprise Data Platform} \\
    \textit{Bell Mobility}\\
    Montreal, QC, Canada \\
    yuecai.zhu@bell.ca}
}

\maketitle

\begin{abstract}
Modern Software Engineering era is moving fast with the assistance of artificial intelligence (AI), especially Large Language Models (LLM). 
Researchers have already started automating many parts of the software development workflow. 
Requirements Engineering (RE) is a crucial phase that begins the software development cycle through multiple discussions on a proposed scope of work documented in different forms. 
RE phase ends with a list of user-stories for each unit task identified through discussions and usually these are created and tracked on a project management tool such as Jira, AzurDev etc. 
In this research we developed a tool ``\textbf{\textit{GeneUS}}'' using GPT-4.0 to automatically create user stories from requirements document which is the outcome of the RE phase.
The output is provided in JSON format leaving the possibilities open for downstream integration to the popular project management tools.
Analyzing requirements documents takes significant effort and multiple meetings with stakeholders. We believe, automating this process will certainly reduce additional load off the software engineers, and increase the productivity since they will be able to utilize their time on other prioritized tasks. 
\end{abstract}



\section{Introduction}
A ``user story''~\cite{lucassen2016use} is commonly used in the Agile software development process. 
It is a description of a unit task that contains the overall description of a particular task, including what to develop, why users/stakeholders need it, and how it should be developed. 
In addition, a user story typically includes functional and non-functional constraints, acceptance criteria, a clear definition of when a task can be marked as ``Done'', and often test case and coverage specification. 

Traditionally, software engineers create a requirements document during the RE phase after several back-and-forth meetings with the stakeholders. 
Engineers then distill these requirements into individual tasks by creating and adding user stories into the project management system. 
This process is heavily effort-consuming and requires a large amount of time from the developers. 
Senior developers, team leads, QA leads, project managers, and scrum masters mostly remains busy interpreting client’s statements into specific unit tasks that are easy for developers to understand. 

Leveraging large language models(LLMs) to automate software engineering processes and development activities is becoming extremely popular and evolving rapidly in both academia and industry. 
To have LLMs automatically complete tasks, we need to develop a mechanism to send appropriate instructions to the model. 
Such instructions are called prompts. The engineering of prompts is commonly known as prompt engineering~\cite{wang2023review}, which is a fundamental methodology in the field of responsive AI. 
As the development of LLMs progresses, the importance of prompt engineering becomes increasingly evident. 
Designing suitable prompts for specific tasks has emerged as a meaningful research direction.
Our study involves extensive work on prompt engineering to generate user stories from the high-level semi-detailed requirements specifications.
This research is a unique contribution to the automation of software engineering processes using LLMs. 
As per our knowledge no such study has been conducted yet to generate user stories with necessary functional and test specifications automatically using LLMs. 

Pre-trained LLM, including ChatGPT~\cite{white2023chatgpt} and Google Palm~\cite{googlepalm}, have not been developed as intelligent as a human.
Simply asking the LLM to provide user stories and test cases with the requirements document does not generate a desirable and useful outcome. 
To overcome this challenge, we propose a prompting technique: Refine and Thought (RaT), which is a specialized version of Chain of Thought (CoT) prompting~\cite{yao2024tree}. 
RaT prompting instructs the LLM to filter out meaningless tokens and refine redundant information from the input in the thought chain. 
RaT optimizes pre-trained LLM's performance in handling redundant information and meaningless tokens and significantly improves the generated user stories and test cases in our application.

To validate the result, we develop the RUST (Readability Understandability, Specifiability, Technical-aspects) survey questionnaire and send it to 50 developers of various backgrounds having experience in creating, reading, and tracking user stories in Agile software development environment. 
The quantitative analysis of the survey participants' feedback shows that the performance of our approach is highly acceptable with 5\% missing technical details, and 1\% of ambiguous task-description, and 0.5\% duplicates.

\subsection{Contributions}
The ultimate outcome of this research is a tool ``\textbf{\textit{GeneUS}}'' that takes a requirements document from users and delivers the detailed user-stories specifying the fundamental attributes~\cite{jira-user-stories}:
\begin{itemize}
    \item Who: The user involved in the task/feature as an actor.
    \item What: Functionality that needs to be achieved.
    \item Why: Purpose of this task/feature.
\end{itemize}
The user stories should also include the following details for the developers:
\begin{itemize}
    \item A definition of done.
    \item Functional and non-functional constraints.
    \item Test specifications.
\end{itemize} 

The ``\textbf{\textit{GeneUS}}'' provides the output in JSON format, leaving it open for integration with existing project management systems where these auto-generated user stories can be used as input to create design diagrams in software design platforms or project management systems.

As per our knowledge, it is the first attempt to automate the RE phase with LLM, which leads to a new SE research domain towards automating the Agile development process as we call \textit{AutoAgile} which aims to automate the entire Agile development process end-to-end with the help of pre-trained LLMs, knowledge embedding, and prompt-engineering. 
In this paper, we report only the attempt to automate the user story generation. 

Our study aimed to answer the two research questions as listed below:

\noindent
\textit{\textbf{RQ1: Can we make LLM produce improved user story specification by applying Refinement and Thought (RaT) into the prompts?}}\\
    Matching the user stories with the requirements document is a challenging task, so we used CoT~\cite{yao2024tree} to do that. However, understanding how thoughts are broken down in CoT is still unclear. Certain difficulties in making automated user stories need additional attention. This pushes us to improve CoT to RaT.

\noindent
\textit{\textbf{RQ2: Is the generated output accepted by a wide variety of software engineering professionals?}}\\
    We wanted to understand if the automatically generated user stories are meaningful and can communicate with human experts. We applied a qualitative analysis on the survey responses to answer this research question.

\section{Background and Related Works}
In Agile development practice developers usually starts a project development with tasks backlog even before the development sprints are kicked off. 
Multiple meeting sessions are required by the development team where mostly the senior members of the team spend hours to interpret customer's statements into unit user-stories or tasks. 
User stories are the key. 

\subsection{The Agile Practice}
In today's software engineering world, the Agile~\cite{cohn2005agile} method is the most popular software development process.
In any software development process requirements engineering is the first step where clients/customer/stakeholders provide a document explaining their needs.
Usually this document explains the type of application, required functionalities, preferred user-interactions and look-and-feel etc. 
Often such details are not available in the very initial project description from the client/customer side. 
However, developers during the Requirements Engineering (RE) stage detail the customer's needs specifying the functionalities, constraints, and other technical or non-technical dependencies, and prepare a Requirement Analysis Document (RAD). 

In the Agile development process, developers need to create a backlog of unit tasks from RAD so that they can create development sprints, estimate milestones, and assign these tasks to the developers.
Developers can get clear instructions on what to develop, why this is needed, how to develop/implement, and how to test and validate the implementation.
In modern Agile development practice, these tasks are written in a format which is commonly known as user stories~\cite{cohn2004user}. 

\subsubsection{User Stories}
In software engineering, a user story is a concise description of a software feature told from the perspective of the end user. 
It defines the functionality or requirement that the user needs and is typically written in non-technical language. 

According to Locassen et al.~\cite{lucassen2016improving}, the qualities of a good user story can be grouped as syntactic, semantic, and pragmatic. 
Syntactic qualities focus on how well-formed a user story is. Semantic qualities focus on whether a user story is conceptually sound, unambiguous, and conflict-free.
Pragmatic qualities focus on a user story's uniqueness, independence, and completeness. 
Other studies~\cite{kuhail2022user} have discussed similar quality attributes for a good user story. 
The common quality attributes are mostly agreed by almost all researchers in the literature and we considered the state-of-the-art quality attributes of user stories as addressed by researchers as well as widely practiced in the industries.
We consider the following attributes as the quality metrics of a user story:
\begin{enumerate}
    \item Readability - User stories should be clear enough for the development team to estimate the effort required for implementation. 
    \item Understandability
        \begin{itemize}
            \item Each user story should deliver value to the end-user or customer. It should address a specific need or provide a solution to a problem.
            \item User stories should be open to discussion and negotiation between the development team and stakeholders to ensure clarity and alignment with project goals.
            \item User stories should be written in such a way that allows clear and measurable acceptance criteria to be defined, enabling easy testing and validation of the implemented functionality.
        \end{itemize}
    \item Specifyability
        \begin{itemize}
            \item User stories should focus on a specific aspect of functionality.
            \item User stories should be specific and small enough to be completed within a single iteration or sprint.
            \item Each user story should be self-contained and independent, allowing it to be implemented and tested without reliance on other stories.
        \end{itemize} 
    \item Technical Aspects
        \begin{itemize}
            \item Each user story should deliver technical details to the developers. 
            \item It should provide technical insight about what to address for the specific need of the problem specified.
        \end{itemize}
\end{enumerate}

We have designed our ``\textbf{\textit{GeneUS}}'' adhering to these criteria to effectively communicate requirements and guide the development process in agile software development.

\subsection{Pre-trained Generative Large Language Model}
Recent advancements in natural language processing(NLP) have shifted from designing task-specific language models to task-agnostic language models~\cite {brown2020language}. 
We are approaching artificial general intelligence (AGI) as the capacity of such task-agnostic language models continues to improve. 
The great success of transformer~\cite{vaswani2017attention} leads to the invention of different pre-trained generative large language models(pre-trained LLM), such as chatGPT~\cite{radford2018improving, radford2019language} and Google Palm~\cite{chowdhery2023palm}. 
These models are pre-trained on vast amounts of text data. 
They have demonstrated exceptional performance in complicated NLP tasks, such as language translation, text summarization, document classification, and logical reasoning in question-answering~\cite{liu2023summary}. 
They have been used in various domains, including education~\cite{lo2023impact}, medication~\cite{hsu2023examining} and legal services~\cite{biswas2023role}. 
Explorations using pre-trained LLMs in software engineering are emerging as a trending research in today's world.

White et al.~\cite{white2023chatgpt} present prompt pattern techniques for software engineering tasks such as improving code quality, refactoring, requirements elicitation, and software design using ChatGPT.
Their study had two key contributions to the field of using LLMs in software engineering. 
Firstly, they presented a catalogue categorizing software engineering patterns based on the problems they address. Secondly, they investigated various prompt patterns aimed at enhancing requirements elicitation, rapid prototyping, code quality, deployment, and testing processes.
We applied a similar pattern while writing prompts to generate user stories in our study and further improved the prompting techniques specific to generating user stories.

Liu et al.~\cite{liu2024your} evaluated the performance of pre-trained LLM in code generation.
Poesia et al.~\cite{poesia2022synchromesh} proposed a reliability-improved model for automated code generation using pre-trained models. 
Ugare et al.~\cite{ugare2024improving} considered grammar to improve the code generation performance by LLMs.
Ahmad et al.\cite{ahmad2023towards} propose collaborative software architecture design with pre-trained LLM. 
Tofano et al.~\cite{tufano2021towards, tufano2022using} study how to automate the code review process using pre-trained LLMs. 
Although many of the software engineering activities have been automated or undergoing research for automation, the automated user story creation from requirements documents has not yet been explored at an implementation level. 
We take the opportunity to become the first to provide a tool that automatically distills the requirements documentation into unit user-stories with necessary specifications, and test coverage which can be automatically integrated into the third-party project management tools for modern Agile software project management. 

\subsection{Prompt Engineering}
\label{prompt_engineering}
Prompt engineering is a newly developed important technique to enhance the performance of pre-trained LLMs. We propose a new prompting technique \textbf{\textit{RaT}}, specially designed to optimize LLM in processing input with redundant information and unrecognizable tokens. 
RaT is a special version of CoT. We will formalize some foundational prompt engineering techniques in this section and introduce RaT in section \ref{Method}. 

Let $p_\theta$ denote a pre-trained LLM with parameters $\theta$, and lowercase letters $x,y,z$ denote language sequences. Then, the scenario in which we asked the question $x$ to the LLM $p_\theta$ and the reply is $y$ can be written as:
\begin{equation}
y \sim p_\theta(x)
\end{equation}

The symbol $\sim$ indicates the random nature of LLM. Depending on the temperature setting of the pre-trained LLM, it is possible that we ask the same questions to the pre-trained LLM multiple times and receive different answers. 
Therefore, $p_\theta$ is a probability distribution of the reply $y$ given the input $x$. 
Then a $n$ step interaction with the LLM can be written as a product of the conditional probability:
\begin{equation}
\label{llm-equa}
    y \sim \Pi_{i=1}^{n} p(x_n|x_1,\dots x_{n-1})
\end{equation}

\subsubsection{Input-Output(IO) Prompting}
Input-Output(IO) prompting simply wraps the input $x$ with task instructions as in zero-shot prompting\cite{radford2019language} or with thought process examples as in few-shot prompting\cite{brown2020language}. IO prompting is the basic building block of more complicated prompting techniques. Let $prompt_{IO}(x)$ represents the IO prompting process wrapping the input $x$, then our interaction with the LLM is presented as:
\begin{equation}
    y \sim p_\theta(y|prompt_{IO}(x))
\end{equation}

For simplicity, we follow the same practice in Yao et al. \cite{yao2024tree} to denote IO prompting as $y \sim p_\theta^{IO}(x)$. 

\subsubsection{Chain of Thought(CoT) Prompting}
CoT\cite{wei2022chain} was proposed to address the cases where the thought process to obtain $y$ from $x$ is complicated. For example, when $x$ is a math problem and $y$ needs further reasoning depending on the answer of $x$. CoT decomposes the thought process from input $x$ to output $y$ into multiple meaningful intermediate steps $z_1, \dots, z_n$. Such that each thought in the chain $z_i \sim p_\theta(z_i|x,z_1,\dots,z_{i-1})$ is sampled sequentially. Then the final output $y \sim p_\theta(y|prompt_{CoT}(x,z_1,\dots,z_n))$, or simply as $y \sim p_\theta^{CoT}(x)$.

In our research, the mapping between the user stories and the requirements document is not trivial, so we applied CoT to optimize the result. However, the decomposition of thought remains ambiguous in CoT\cite{yao2024tree}. Some particular challenges presented in automated user story generation require special treatment. This motivates us to optimize CoT to RaT, which will be detailed in section \ref{Method}.

\section{Methodology}
\label{Method}
\begin{figure}[ht!]
\includegraphics[width=\columnwidth]{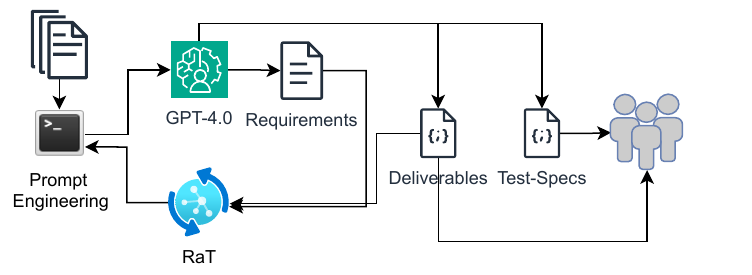}
\caption{Methodology}
\label{fig:methodology}
\end{figure}

Our methodology begins with prompt engineering, specifically for giving instructions to an LLM to understand the software requirements documents. 
The requirements document is passed through the prompts into the LLM which extracts the functional and nonfunctional requirements. 
The requirements then passed into RaT for refinement and resent to the LLM in the second short which generates the deliverables with a clear definition of done, task specifications, technical details and acceptance criteria. 
Further, the third shot of prompts are sent to the LLM to generate the test coverage specifications. 
Figure~\ref{fig:methodology} shows the methodology at a glance. 

As introduced in section~\ref{prompt_engineering}, prompt engineering is a technique to enhance the performance of the pre-trained LLM. Especially when the reasoning process to find the ultimate answer $y$ for the original input question $x$ is not straightforward. In this case, the technique CoT\cite{wei2022chain} is often applied. However, the decomposition of the thought process in CoT remains ambiguous. Besides the complicated reasoning process,  we also face the challenge of less desirable results caused by redundant or meaningless tokens from the input. This motivates us to develop RaT prompting, a variant of CoT. 
\subsection{Refine and Thought(RaT) Prompting}
The requirements document must be transformed into pure text before we can ask the pre-trained LLM to process it.
Requirements documents usually contain pictures to interpret graphical designs or flow diagrams to present the application's logical decision-making process. These non-text elements become meaningless symbols during the pure text extraction from the input document. These unrecognizable symbols negatively impact the quality of the LLM's output. To overcome this challenge, we improve CoT prompting to RaT prompting. 

A RaT prompting block is a two-step CoT that includes the refinement and thought steps. As an example, in the first step of our application to process the input requirement document, 
we first ask the LLM to refine the text with unrecognizable symbols or words in the refinement step. In the thought step, we ask the LLM the actual question with the refined text from the refinement step as the provided context. RaT can be expressed as:
\begin{equation}
    y \sim p_\theta(y|prompt_{thought}(z))p_\theta(z|prompt_{refine}(x))
\end{equation}
Which can also be written as the following for simplicity:
\begin{equation}
    y \sim p_\theta^{RaT}(x) = p_\theta^{thought}(p_\theta^{refine}(x))
\end{equation}

As shown in algorithm~\ref{algo}, multiple RaT prompting blocks can be concatenated together to produce a longer CoT flow in a more sophisticated manner.

As discussed in section~\ref{hallucination}, RaT improved the consistency of the results obtained from the prompting interaction with LLM and reduced GenAI hallucinations. 

%The proposed technique RaT helps improve the result from the pre-trained LLM.

\subsection{Automated User-Stories and Test Cases Generation}
Transforming requirement documents to user stories and test case specifications has four steps: document text extraction, requirements extraction, user stories generation, and test case generation. Each of the three latter steps requires a RaT prompting block. Three RaT prompting blocks form the complete chain of thought. Algorithm~\ref{algo} shows the complete process. We also provide an online REST API for researchers to test the proposed application.

\begin{algorithm}
\caption{Automated user-stories and test-cases generator}
\SetAlgoLined
\SetKwData{Left}{left}\SetKwData{This}{this}\SetKwData{Up}{up}
\SetKwFunction{Union}{Union}\SetKwFunction{FindCompress}{FindCompress}
\SetKwInOut{Input}{input}\SetKwInOut{Output}{output}
\Input{$document$}
\Output{$user\_stories, test\_cases$}
initialization\;
$x_1\leftarrow$ "generate user stories for the input text";\\
$x_2\leftarrow$ "generate test cases for the input user stories.";\\
$x_3\leftarrow$ "add deliverables to each user story.";\\

$text\leftarrow$ $textExtract(document)$ ;\\
$requirements\leftarrow p_\theta^{RaT}(text, x_1)$;\\
$test\_cases\leftarrow p_\theta^{RaT}(requirements, x_2)$;\\
$user\_stories \leftarrow p_\theta^{RaT}(requirements, x_3)$;\\
\label{algo}
\end{algorithm}

We test the GeneUS with seven different sample RE documents to generate user stories, their deliverables, and test case specifications for each requirement. 
A sample output snippet is attached in the Appendix A. 
We also survey developers to evaluate the quality of the output. 

\subsection{Data}
We collected seven mid-sized RE documents. 
Six of them were collected from the case studies available in the popular SE text-book ``\textit{Software Engineering, 10th Edition}'' by Ian Sommerville~\cite{sommerville} which is taught in many universities at both Undergraduate and Graduate levels.
One of the RE documents comes from an event management software project in a local software development company.
Our tool can process only text inputs at this point, therefore, we removed all diagrams, and UI mockups from the RE documents before we pass it to our tool to avoid any ambiguity for the LLM. 

\subsection{Survey}
We conducted a survey among 50 developers from various backgrounds to validate the output of ``\textbf{\textit{GeneUS}}''.
We sent email invitations to 76 prospective participants who are working in five different software development industries, and 2 universities located in the USA and Canada.

\begin{table}[ht!]
    \centering
    \caption{RUST Survey Questionnaire}
    \begin{tabular}{|l|p{7.5cm}|}
        \hline
        \multirow{5}{*}{R} & Q1. How readable do you think these user-stories are? \\
        %\hline
         & Q2. How grammatically correct are the user stories? \\
        %\hline
         & Q3. How do you rate the spelling accuracy in the output text? \\
        %\hline
         & Q4. Are the descriptions too verbal and contain redundant information? \\
        %\hline
         & Q5. How readable do you think these test cases are? \\
        \hline
        \multirow{5}{*}{U} & Q1. How would you rate the semantic meanings accuracy in the user stories? \\
        %\hline
         & Q2. How would you rate the clarity of the language used in the user stories? \\
        %\hline
         & Q3. How comprehensible were the user stories provided to you? \\
        %\hline
         & Q4. Did you encounter any difficulties in understanding the content of the user stories? \\
        %\hline
         & Q5. Did you feel confident in your understanding of the objectives and requirements outlined in the user stories? \\
        \hline
        \multirow{5}{*}{S} & Q1. Did you find the user stories contained sufficiently specific information to estimate the effort required for implementation? \\
        %\hline
         & Q2. How well-specified were the acceptance criteria outlined in the user stories? \\
        %\hline
         & Q3. Were there any essential details missing from the user stories that you believe would be crucial for effective implementation? \\
        %\hline
         & Q4. How well-specified were the test scenarios outlined in the test cases? \\
        %\hline
         & Q5. How confident are you to break the stories down to multiple sub-tasks given the specifications? \\
        \hline
        \multirow{5}{*}{T} & Q1. How well do the given test cases perform the technical requirements and the corner cases? \\
        %\hline
         & Q2. Do you think the design of the deliverables complies with the software engineering best practices? \\
        %\hline
         & Q3. The user stories interpret the user/stakeholders’ statements into technical language including technical details and insights? \\
        %\hline
         & Q4. The user stories adequately outline the technical requirements and constraints for the given tasks. \\
        %\hline
         & Q5. Were there any ambiguities or inconsistencies in the technical details provided within the user stories? \\
        \hline
    \end{tabular}
    \label{tab:rust}
    \label{tab:my_label}
\end{table}

The survey questionnaire focuses on the standard quality attributes of a good user story and allows participants to provide their feedback guided by RUST.  
Table~\ref{tab:rust} shows the survey questions we sent to the participants. 
The first column indicates the categories of the questions where \textit{R}, \textit{U}, \textit{S}, and \textit{T} represents \textit{Readability}, \textit{Understandability}, \textit{Specifiability}, and \textit{Technical aspects} respectively.
The survey response is discussed in details in Section~\ref{result}. 

\section{Results and Discussion}
\label{result}
\textbf{\textit{GeneUS}} provides a detailed JSON string containing the unit tasks from the RE documents. 
Appendix A shows a portion of the JSON output for one of the user stories generated from the \textit{Mental Health Support System}.
In this study the biggest challenge was Hallucination~\cite{xu2024hallucination}.
Hallucinations are not bugs~\cite{yao2023llm}, LLMs suffer from hallucinations (Bang et al., 2023; Lee et al., 2018) which means
LLMs lie and fabricate non-existent facts or inappropriate information.

\subsection{LLM Hallucinations Under Complex Context}
\label{hallucination}
As introduced in section~\ref{Method}, one of the challenges we faced in the development of GenUS is that the LLM either gives incomplete information or wrong information in the result. 
Moreover, the LLM answers are inconsistent if we ask the same questions multiple times. 
This is called LLM hallucination. 
Hallucination is well explored in Ji et al.~\cite{ji2023survey}. 
It's very challenging to detect and control hallucinations in real-world LLM applications. 
The cause is rooted in the very nature of LLM, that it's a probabilistic model. 
As described in equation~\ref{llm-equa}, LLM predicts the next possible word based on the previous input sequence. 
Therefore, the longer the input word sequences are, the higher the possibility hallucinations in the LLM's answers. 

The challenge of LLM hallucinations is even worse in RAD. 
Software requirement documents are usually very long and complex. 
Asking the LLM to extract specific information from such long and complex context often resulted in answers with hallucinations. 
For example, one of the software requirement documents we tested has the following text:

\begin{mdframed}
\textit{Diabetics measure their blood sugar levels using an external meter and then calculate the dose of insulin that they should inject. The problem with this treatment is that the level of insulin required does not just depend on the blood glucose level but also on the time of the last insulin injection. This can lead to very low levels of blood glucose (if there is too much insulin) or very high levels of blood sugar (if there is too little insulin). Low blood glucose is, in the short term, a more serious condition as it can result in temporary brain malfunctioning and, ultimately, unconsciousness and death. In the long term, however, continual high levels of blood glucose can lead to eye damage, kidney damage, and heart problems. Current advances in developing miniaturized sensors have meant that it is now possible to develop automated insulin delivery systems. These systems monitor blood sugar levels and deliver an appropriate dose of insulin when required.}
\end{mdframed}

The text explains the problem with the current practice of insulin dosage calculation, which is that it requires not only the estimation of blood glucose level but also the time of the last insulin injection. The system addresses this problem by monitoring patients' blood sugar levels in real time and calculating the dose of insulin to be delivered only by blood sugar level.

However, since this context is long and complex, the LLM we used in the experiment (``\textit{gpt-4-1106-preview}''), wrongly captures the requirements if we ask the LLM to extract the list of requirements from the original document (see below).

\begin{lstlisting}
The system shall collect information from a sensor to monitor blood sugar levels.
The system shall calculate the required dose of insulin based on the blood sugar level and time since the last insulin injection.
The system shall control the pump to deliver the calculated dose of insulin to the user.
The system shall ensure that the pump delivers the correct amount of insulin in response to the controller's signals.
The system shall be able to deliver insulin in units, with each unit corresponding to a single pulse from the controller.
The system shall be available to deliver insulin when required to the user.
The system shall perform reliably to maintain the user's blood sugar levels within a safe range.
The system shall be designed and implemented to ensure the safety of the user's health by preventing incorrect insulin delivery.
\end{lstlisting}

Since the driver of LLM hallucinations is the length and complexity of the input context, we address this problem by proposing RaT to shorten the input context and make it more concise and precise. The following is the output of functional requirements extracted from the input requirement document refined with the help of RaT:
\begin{lstlisting}
The system must collect real-time data from a sensor to monitor blood sugar levels.
The software must calculate the required amount of insulin based on the current blood sugar level.
The system should be able to control the insulin pump to deliver the calculated dose of insulin to the user.
The software must have mechanisms in place to ensure the correct amount of insulin is delivered accurately.
The system should be able to send signals to the pump promptly when insulin delivery is required.
The software must have fail-safe mechanisms to prevent over or under delivery of insulin.
The system should provide alerts or notifications to the user in case of any issues or failures in the insulin delivery process.
The software must be designed to operate reliably to ensure the health and safety of the user.
The system should have backup systems in place to ensure continuous operation in case of any failures.
\end{lstlisting}

The RaT-based prompting captures the requirement much better than the one not using RaT. 
The output from RaT captures the requirement to use a sensor to monitor the patient's blood sugar level and deliver the insulin needed without considering the time since the last injection, while the output without RaT mistakenly takes the time since the last injection as one of the conditions. 

We ran the same experiment for each document in the test 10 times. 
Even though the outcomes for the 10 runs are not entirely same in wording, the requirements captured are the same. 
RaT prompting also improve consistency of the LLM. 
% Answer RQ1
\begin{mdframed}
\textbf{\textit{Answer to RQ1: We are able to make LLM produce improved user story specification by applying Refinement and Thought (RaT).}}
\end{mdframed}

\subsection{RUST Survey Result}
We conducted testing on our tool using seven Requirements Engineering (RE) documents. 
Subsequently, we distributed the input and output JSON files to 50 developers, who evaluated the generated user stories based on their expertise and familiarity with software requirements engineering. 
Notably, the participants hailed from both the software industry and various universities across two distinct countries. 
Despite their diverse backgrounds, all participants shared a common foundation in software engineering, possessing a minimum of one year of professional experience in the field. 
On average, participants had four years of experience in software requirements analysis and user story creation, with a maximum of ten years and a standard deviation of 1.86. 

The survey questionnaire was structured into four sections. 
Participants began by reviewing the requirements document and subsequently assessed the user stories generated. 
Responses to each question were recorded on a scale of 1 to 5, with 1 denoting the poorest feedback and 5 representing the most positive response. 
Notably, while the median feedback score for two of the Readability questions was 3, the median score for the RUST questions was 4 overall. 
Additionally, some participants identified certain keywords or phrases as spelling errors when utilizing their chosen spell-checker tool. 
This observation may stem from participants' perception that certain user stories could have been consolidated within others. 
As user stories ideally should be as unit as possible, in alignment with the model's instructions, the LLM diligently adhered to this guidance, striving to generate user stories as unit as possible.

\begin{figure}[ht!]
\centering
\includegraphics[width=\columnwidth]{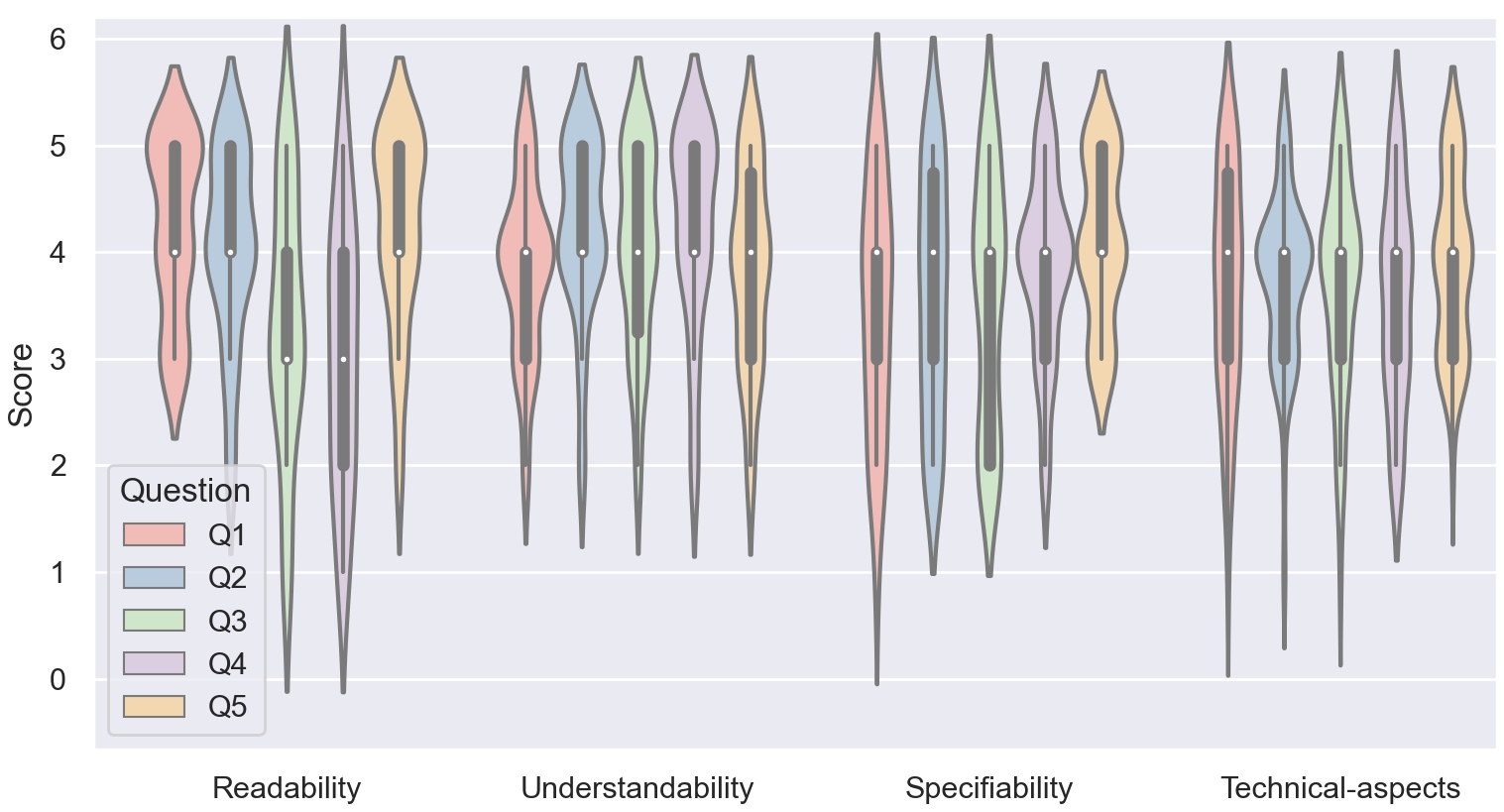}
\caption{Survey Results for Each Question}
\label{fig:rush-scores}
\end{figure}

The bean plot depicted in Figure~\ref{fig:rush-scores} illustrates the scores for each question within the RUST survey's respective groups. 
In particular, Question 5 of the Readability section solicits participants' overall assessments of readability. 
The median score, representing a collective evaluation, stands at 4, with the majority of feedback scores surpassing this value. 

Although the Understandability, Specifiability, and Technical Aspects groups consistently received a median feedback score of 4, several responses fell below this threshold. 
Specifically, for Understandability, Question 1 elicited feedback below 4. 
Similarly, for Specifiability, Questions 1, 3, and 4 received feedback below 4, indicating concerns regarding semantic accuracy, sufficiency of specific information, essential details, and test case specifications.

Question 1 of the Understandability section focuses on the accuracy of semantic meaning within the user stories. 
In the Specifiability section, Questions 1, 3, and 4 assess the adequacy of specific information, essential details, and test case specifications, respectively. 
Although the median scores for these questions are above neutral, a significant portion of respondents expressed dissatisfaction with the level of detail provided in the specifications.

Likewise, feedback below 4 was prevalent for all technical-aspects questions except Question 1. 
This suggests that some participants harbored doubts regarding the complete alignment of deliverables with software engineering standards, the clarity of interpretation from high-level descriptions to technical tasks, and the unambiguous nature of technical details.

\begin{figure}[ht!]
\centering
\includegraphics[width=\columnwidth]{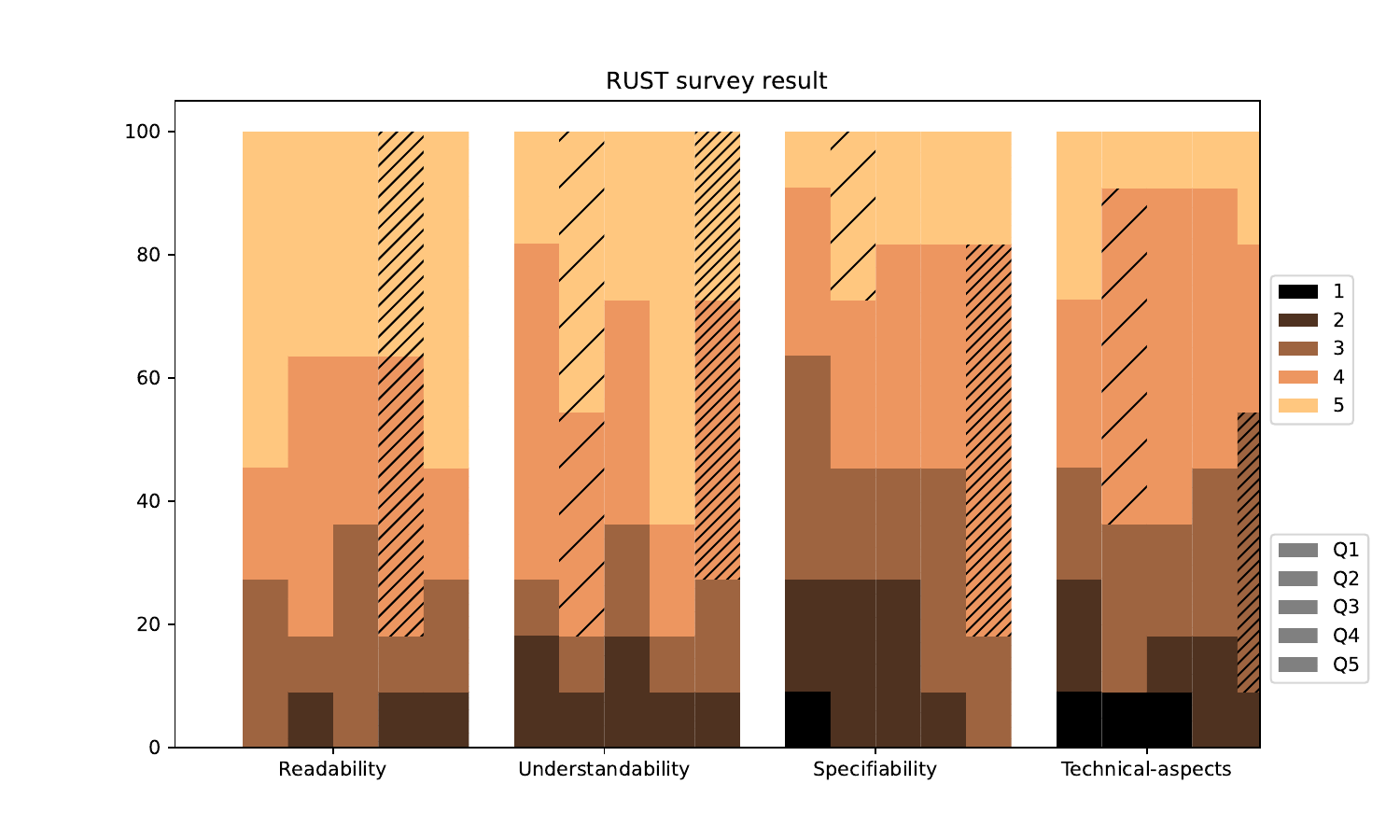}
\caption{Survey Result Heat-map for Each Group of Assessment}
\label{fig:rust-heatmap}
\end{figure}

Figure~\ref{fig:rust-heatmap} presents an alternate visualization of the survey outcomes in the form of a heatmap. 
Darker regions indicate lower scores, while brighter areas signify higher scores. 
Notably, a significant portion of the heatmap appears bright or semi-bright, with only minimal dark areas. 
Darker zones are slightly more prominent within the Specifiability and Technical Aspects categories compared to Readability and Understandability. 
This observation suggests that the LLM demonstrates stronger performance in natural language tasks but there are areas of improvements in context-specific endeavors.

\begin{mdframed}
    \textbf{\textit{Answer to RQ2: The generated user stories were rated as ``Good'' (4 out of 5) and accepted by the majority of the SE experts participated in the RUST survey. However, the user stories have room for improvements, especially in the Specifiability and Technical-aspect categories.}}
\end{mdframed}

\section{Conclusion}
User stories serve as foundational components within Agile software engineering practices. 
They play a pivotal role in project planning, estimation, distribution, and progress tracking across the software development lifecycle. 
A well-crafted user story facilitates effective communication among developers, project planners, architects, quality controllers, and stakeholders involved in the software project. 
In modern software development practices, the utilization of project management tools such as Jira or Azure DevOps for creating and managing user stories has become indispensable. 
However, transforming high-level requirements engineering documents into tangible user stories often consumes considerable time, primarily from senior team members. 
Our objective in automating the user story creation process is to alleviate this burden from developers, enabling them to allocate their time to other priority tasks and thereby enhance productivity.

In this context, we introduce a tool named \textbf{\textit{GeneUS}} designed to \textbf{\textit{Gene}}rate \textbf{\textit{U}}ser \textbf{\textit{S}}tories from requirements documents. 
The tool outputs data in JSON format, facilitating seamless integration with popular project management platforms. 
To enhance the precision of user story creation and mitigate the risk of hallucinations within the Language Learning Model (LLM), we applied Refine and Thought (RaT) prompting strategy. 
RaT shortens the input context, rendering it more concise and precise, as the length and complexity of input data can influence LLM's tendency to hallucinate.

Upon manual comparison of automatically generated user stories, we discovered that responses without RaT occasionally omitted crucial context and were susceptible to hallucinations. 
Engaging 50 Software Engineering (SE) practitioners from diverse backgrounds in Canada and the United States, we collected feedback using the RUST questionnaire. 
Results indicate that the overall quality of the generated user stories falls just shy of excellence, with a median score of 4 out of 5. 
Despite commendable scores, areas such as Specifiability and Technical Aspects show potential for improvement.

Moving forward, we plan to enhance the tool's performance by subjecting it to additional requirements engineering documents sourced from a broad spectrum of software projects. 
Our tool will be made publicly accessible to gather requirements engineering documents from global sources, facilitating iterative testing and refinement of prompts. 
Additionally, we aim to integrate domain-specific knowledge and leverage knowledge embedding techniques to effectively address issues related to hallucinations. 

\bibliographystyle{IEEEtran}
\bibliography{paper}

\onecolumn
\appendices

\section{Sample Input}
Following is a sample input to the GeneUS which is a high level requirement specification document of ``\textit{MentCare Mental Health Support System}'' collected from the Software Engineering text book by Ian Sommerville~\cite{sommerville}.

\begin{mdframed}[backgroundcolor=black!10,rightline=false,leftline=false]

\begin{center}
Patient Information System - MentCare\\
------------------------------------------------
\end{center}

A patient information system to support mental health care is a medical information system that maintains information about patients suffering from mental health problems and the treatments that they have received. Most mental health patients do not require dedicated hospital treatment but need to attend specialist clinics regularly where they can meet a doctor who has detailed knowledge of their problems. To make it easier for patients to attend, these clinics are not just run in hospitals. They may also be held in local medical practices or community centers.

The MHC-PMS (Mental Health Care-Patient Management System) is an information system that is intended for use in clinics. It makes use of a centralized database of patient information but has also been designed to run on a PC, so that it may be accessed and used from sites that do not have secure network connectivity. When the local systems have secure network access, they use patient information in the database but they can download and use local copies of patient records when they are disconnected. The system is not a complete medical records system so does not maintain information about other medical conditions. However, it may interact and exchange data with other clinical information systems. Figure 1.6 illustrates the organization of the MHC-PMS.

The MHC-PMS has two overall goals:
1. To generate management information that allows health service managers to assess performance against local and government targets.
2. To provide medical staff with timely information to support the treatment of patients.

The nature of mental health problems is such that patients are often disorganized so may miss appointments, deliberately or accidentally lose prescriptions and medication, forget instructions, and make unreasonable demands on medical staff. They may drop in on clinics unexpectedly. In a minority of cases, they may be a danger to themselves or to other people. They may regularly change address or may be homeless on a longterm or shortterm basis. Where patients are dangerous, they may need to be 'sectioned' confined to a secure hospital for treatment and observation.

Users of the system include clinical staff such as doctors, nurses, and health visitors (nurses who visit people at home to check on their treatment). Nonmedical users include receptionists who make appointments, medical records staff who maintain the records system, and administrative staff who generate reports.

The system is used to record information about patients (name, address, age, next of kin, etc.), consultations (date, doctor seen, subjective impressions of the patient, etc.), conditions, and treatments. Reports are generated at regular intervals for medical staff and health authority managers. Typically, reports for medical staff focus on information about individual patients whereas management reports are anonymized and are concerned with conditions, costs of treatment, etc.

The key features of the system are:

1. Individual care management Clinicians can create records for patients, edit the information in the system, view patient history, etc. The system supports data summaries so that doctors who have not previously met a patient can quickly learn about the key problems and treatments that have been prescribed.
2. Patient monitoring The system regularly monitors the records of patients that are involved in treatment and issues warnings if possible problems are detected. Therefore, if a patient has not seen a doctor for some time, a warning may be issued. One of the most important elements of the monitoring system is to keep track of patients who have been sectioned and to ensure that the legally required checks are carried out at the right time.
3. Administrative reporting The system generates monthly management reports showing the number of patients treated at each clinic, the number of patients who have entered and left the care system, number of patients sectioned, the drugs prescribed and their costs, etc.

Two different laws affect the system. These are laws on data protection that govern the confidentiality of personal information and mental health laws that govern the compulsory detention of patients deemed to be a danger to themselves or others. Mental health is unique in this respect as it is the only medical speciality that can recommend the detention of patients against their will. This is subject to very strict legislative safeguards.

One of the aims of the MHC-PMS is to ensure that staff always act in accordance with the law and that their decisions are recorded for judicial review if necessary. As in all medical systems, privacy is a critical system requirement. It is essential that patient information is confidential and is never disclosed to anyone apart from authorized medical staff and the patient themselves. The MHC-PMS is also a safety-critical system. Some mental illnesses cause patients to become suicidal or a danger to other people. Wherever possible, the system should warn medical staff about potentially suicidal or dangerous patients.

The overall design of the system has to take into account privacy and safety requirements. The system must be available when needed otherwise safety may be compromised and it may be impossible to prescribe the correct medication to patients. There is a potential conflict here privacy is easiest to maintain when there is only a single copy of the system data. However, to ensure availability in the event of server failure or when disconnected from a network, multiple copies of the data should be maintained. I discuss the trade-offs between these requirements in later chapters.
\end{mdframed}

\section{Sample Output} Following is a sample output of our tool ``\textit{GeneUS}'' based on the requirements document for the case study ``\textit{MentCare Mental Health Support System}'' from the Software Engineering text book by Ian Sommerville~\cite{sommerville}.
\begin{lstlisting}[language=C]
{
    "User Story": "The system should allow clinicians to create, edit, and view patient records.",
    "Deliverables": {
        "architecture_design": {
            "definition_of_done": "Architecture design document created and reviewed by the
            development team. Design aligns with system requirements and best practices.",
            "criteria": [
                "Document created and stored in a designated location",
                "Design reviewed and approved by development team",
                "Design aligns with system requirements and scalability needs",
                "Design follows best practices and coding standards"
            ]
        },
        "database_schema_design": {
            "definition_of_done": "Database schema designed and implemented for storing patient records.",
            "criteria": [
                "Schema design document created and stored in a designated location",
                "Schema implemented in the database",
                "Schema optimized for efficient data storage and retrieval",
                "Schema reviewed and approved by database administrators"
            ]
        },
        "unit_tests": {
            "definition_of_done": "Unit tests written and executed for patient record management functionality.",
            "criteria": [
                "Unit test cases written to cover all aspects of patient record management",
                "Tests executed and passed successfully",
                "Code coverage of unit tests meets defined threshold",
                "Unit tests integrated into the automated testing framework"
            ]
        },
        "user_training_documentation": {
            "definition_of_done": "Comprehensive documentation created on how to create, edit, and view patient records.",
            "criteria": [
                "Documentation covers all functionalities related to patient record management",
                "Documentation is clear, concise, and easy to understand",
                "Documentation reviewed and approved by stakeholders",
                "Documentation accessible to all users through a centralized location"
            ]
        },
        "production_support_plan": {
        "definition_of_done": "Plan developed for supporting patient record management in a
        production environment.",
            "criteria": [
                "Plan outlines procedures for monitoring and maintaining patient records",
                "Plan includes strategies for handling system failures and data backups",
                "Plan reviewed and approved by operations team",
                "Plan integrated into the overall support and maintenance framework"
            ]
        }
    }
},
{
    ...
    ...
},
\end{lstlisting}

\end{document}